# HIGH PRECISION MICROSCALE 3D MANUFACTURING OF ULTRA LOW EXPANSION GLASS BY FEMTOSECOND SELECTIVE LASER ETCHING

Enrico Casamenti[1,a], Alessandro Bruno[1], Pietro Bernasconi[1], Andrea Lovera[1]

[1] FEMTOprint SA, Via Industria 3, Muzzano, Ticino, Switzerland
[a] **Author to whom correspondence should be addressed**: enrico.casamenti@femtoprint.ch

**Abstract**

Due to its exceptional dimensional stability in harsh thermal conditions and excellent mechanical and optical properties, Corning ultra-low expansion (ULE®) glass is the material of choice in many high-demanding fields such as aerospace, astronomy, and advanced optics. This material has recently attracted renewed attention with the advent of femtosecond laser technology, with a particular focus on the interaction of ultrafast pulses and the material itself. Phenomena like the formation of self-assembled nanogratings and their thermal stability as well as the darkening of laser-affected zones have then been characterized. This paper presents how to exploit femtosecond selective laser etching (SLE) techniques to manufacture truly three-dimensional (3D) components. To demonstrate the micron-scale accuracy and repeatability of this process from the mm- to the cm-size range, various devices are designed and fabricated: fiber hole arrays with different hole densities, sizes, orientations, and shapes; and fiber V-groove arrays. Additionally, a mechanical flexural fiber mount is presented as an example of how multiple functionalities can be monolithically integrated into a single piece of glass through SLE technology. An example of a passive alignment substrate for optical components is also shown. SLE technique represents a new advancement in the field of microscale manufacturing, enabling the scalable production of custom-designed ULE® glass structures with unprecedented precision and complexity, paving the way for the miniaturized integration of highly stable components.

**Introduction**

Since the early 20th century, the development of low thermal expansion glasses has been mostly driven by the stringent requirements set by astronomical telescopes. The material of choice for telescope's mirrors evolved from standard soda lime with a coefficient of thermal expansion (CTE) well above $50·10^{-7}/°C$, to a low expansion borosilicate glass (CTE: $25·10^{-7}/°C$) then to fused silica (CTE: $5·10^{-7}/°C$) and finally, in 1968, to ULE® glass by Corning that reached a CTE of about $0.2·10^{-7}/°C$ [1]. ULE® is an ultra-low expansion titania-silica binary glass with a nominal composition of 7 wt.% $TiO_2$ in $SiO_2$ [2]. This glass shows near-zero thermal expansion over the ambient operating temperature of usual telescopes (CTE down to $30·10^{-9}/°C$ from 5°C to 35°C), no thermal hysteresis, high fatigue resistance, and excellent optical and birefringence properties, which allows extreme reproducibility and homogeneity in the manufacturing [3]. All these properties combined made ULE® one of the best candidates for high-precision applications.

The continuous drive for miniaturization and integration has recently boosted the appeal of this glass within and beyond modern astronomy. For instance, the massive introduction of optical fiber technologies in the design of next-generation telescopes has shifted the interest in highly stable glasses from the mirrors to the detection parts of the systems where fibers capture the light signals collected by the telescope and deliver it to the sensors and cameras to provide data and images with unprecedented accuracy [4, 5]. Furthermore, high-precision laser systems, such as those used in advanced manufacturing and medical procedures, have also benefited from the remarkable properties of ULE® in particular for components that must maintain their opto-mechanical properties under thermal stress. [6 - 8] In laser cutting and welding, for example, optical elements like lenses and mirrors must not deform under high heat to ensure clean, accurate cuts and joins. Similarly, in laser eye surgery, the precision and thermal stability of optical components are critical for successful outcomes [9].

An important contribution to the integration and fabrication of complex glass parts has been given by the developments in the field of femtosecond selective laser etching (SLE) [10 - 13]. This relatively new technology that allows for the manufacturing of nominally any arbitrary three-dimensional shape with dimensions ranging from a few tens of microns to several centimeters in glasses is characterized by its high micrometric accuracy. It relies upon a two-step process: the first one consists of the exposure of the glass substrate to a focused femtosecond laser that defines the outer envelope surface of the object to be produced; the second is a subsequent etching bath (typically based upon HF [10], KOH [14], or NaOH [13]) that removes all excess material and eventually releases the desired part. With this technique, for instance, Nazir et al. developed multiple high-precision mechanical alignment devices [15, 16], that were afterward integrated into a miniaturized optical assembly [17]. Among many other microdevices (e.g. [18 - 21]), a puncturing tool for eye surgery [22], and glass ferrules for 2D fiber arrays [23] were also recently demonstrated.

Our fabrication method relies upon the physical processes that govern the interaction of tightly focused, ultrafast laser pulses with the glasses. Among these, the formation of nanogratings, their thermal stability, and the darkening of laser-affected zones — have been extensively investigated, and described in the literature [24 - 29]. In this paper, we prove that the SLE principles can be successfully applied to the manufacturing of truly 3D elements in ULE®. The results will show that micrometric precision and sub-micron surface roughness can be achieved, opening up countless design opportunities in the most various fields, from astronomy to sensors, to medical instruments, and, more in general, wherever high precision and optomechanical, thermal stability is required.

### Devices' fabrication & characterization

A set of opto-mechanical devices were designed to be fabricated in ULE® according to the FEMTOPRINT® procedures. Building on the work available in the literature about nanogratings formation in ULE® via femtosecond irradiation [24 - 26, 29], a combination of laser parameters that lead to efficient selective laser etching is found and applied.

Multi-fiber optical ferrules

The first group of devices comprises a series of optical multi-fiber aligners that require the highest accuracy (micron or sub-micron) in both the positioning and the diameters of the holes.

The first example consists of a ferrule for standard single-mode optical fibers with a nominal diameter of 125 μm. As shown in Figures 1A-C, the 2 mm-thick device contains an array of 960 vertical holes arranged in a honeycomb pattern with a pitch of 250 µm. It can be appreciated from Figure 1C that the SLE fabrication delivers very uniform, sharp, and clean cuts without burrs or chippings along the edges of the openings. It is worth mentioning that the current pitch of 250 µm is not a limiting factor, as it can be reduced to a few tens of microns, if necessary. The present choice was actually dictated by the presence of conical inlets on the opposite side of the openings to facilitate the insertion of the fibers. Furthermore, as shown by the company's logo in Figure 1A, this technique allows to engrave in-volume markers or fiducials with the same positioning and dimensional accuracy (see below) achievable with the etched features to be used as references for optical alignment purposes.

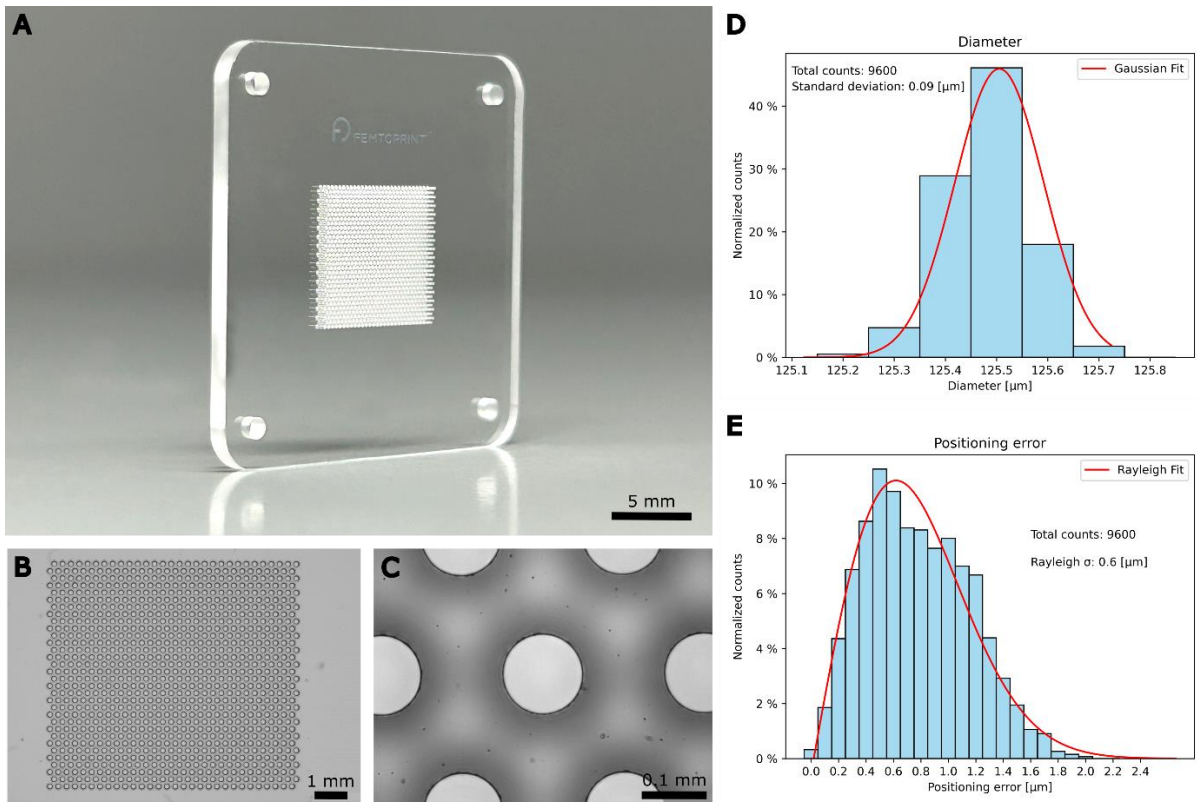

Figure 1. (A-B-C) Increasingly zoomed views of a fiber ferrule with 960 holes of 125 μm in diameter for standard optical fibers positioning in a honeycomb disposition with a pitch of 250 μm. (D) Holes' diameters distribution with a Gaussian fit. (E) Holes positioning error distribution with respect to the nominal grid with a Rayleigh fit.

The finished device is characterized under a cross-calibrated optical digital microscope to investigate the precision and reproducibility of the holes in terms of both diameter and positioning. The results of 10 measurement runs performed on the same sample, each time removing and repositioning it below the microscope, are plotted in the histogram of Figure 1D. The analysis of the diameter's distribution is fitted by a Gaussian curve, which is centered almost exactly at the target diameter of 125.5 µm (0.5 µm is added to the nominal fiber diameter to allow the insertion) and a standard deviation (σ) as low as 0.09 µm. This translates into an achievable tolerance of $\pm 3\sigma = 0.27\ \mu m$ for this specific process and design. The positioning error (Figure 1E) is defined as the distance between the center of each measured hole and its corresponding position on the nominal hexagonal pattern. The data are fitted with a Rayleigh distribution characterized by a scale parameter σ = 0.6 µm, meaning that about 99% of the positioning offsets are smaller than $3\sigma = 1.8\ \mu m$. Note that this accuracy refers to the absolute maximum deviation detectable over 7.5 mm, i.e. the entire length of the array.

The versatility of the SLE manufacturing technology has been exploited to design different layouts of precise fiber aligners, from more conventional V-groove arrays (Figures 2A-B) to more elaborate hole aligners (Figures 2C-E). The former represents a V-groove array with 16 slots nominally spaced by 300 µm and a total groove angle ϑ of 60°. The measurements reveal an actual pitch of 300.0±0.3 µm and ϑ of 60.04°±0.03° in excellent agreement with the nominal values. Note that, besides the shape and position accuracy, SLE offers great flexibility in choosing even non-uniform pitches or arbitrary groove angles ϑ. The surface roughness of the laser-machined sections was also measured interferometrically leading to an average value Sa ~ 200 nm on vertical walls, and a slightly lower value Sa ~ 150 nm on the slanted planes in the grooves. Figure 2C showcases a fiber aligner with 7 converging fibers on a 2 mm-thick substrate. The purpose of the arrangement is to provide a central vertical hole to host a fiber used for illumination surrounded by 6 other converging (tilted) cavities hosting different kinds of fibers devoted to light collection. The configuration of this optical sensor underlines once again the versatility in the design freedom and precision offered by SLE. Figures 2D-E present an example of the multi-functional integration opportunities offered by SLE. Here, the fiber aligner is connected to the surrounding frame via a set of flexures that provide enhanced thermal insulation of the fiber block from the surrounding environment. This design leverages the exceptional thermal stability of ULE® glass, ensuring that the component maintains its performance even under widely varying thermal conditions.

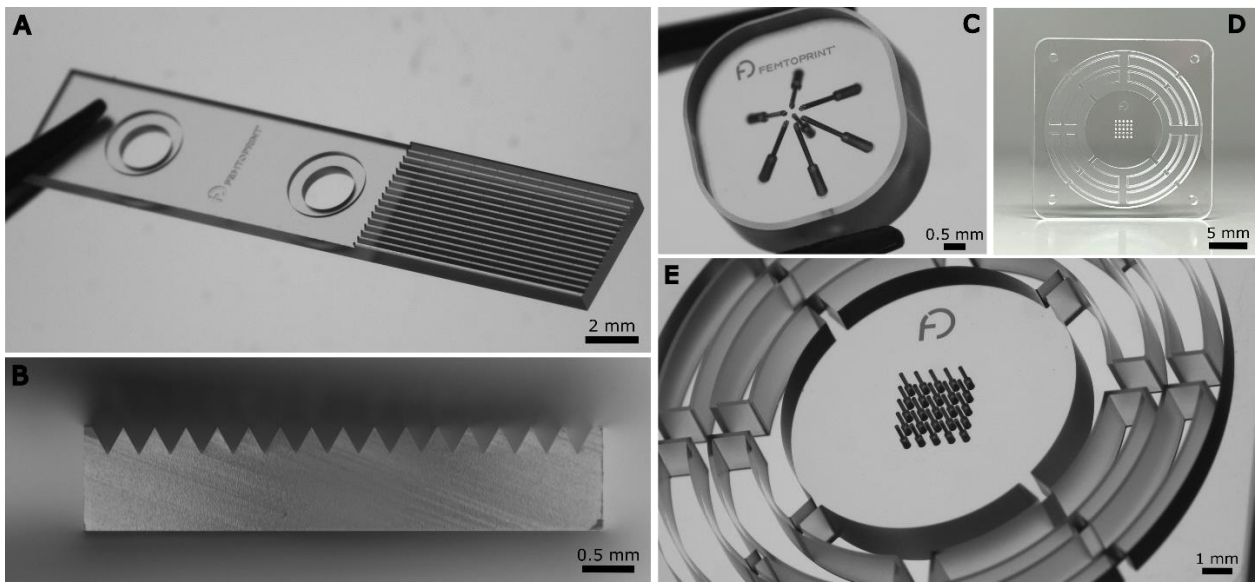

Figure 2. (A) Top view and (B) front view of a 16-channel V-groove array with a pitch of 300 µm and a groove angle ϑ of 60°. (C) Fiber aligner with tilted holes to combine imaging and sensing in a single device. Overview (D) and detail (E) of a fiber aligner designed for improved thermal insulation.

All the above examples highlight how the use of SLE technology allows for the exploitation of ULE® unique glass properties to create complex, multifunctional devices with superior thermal and structural properties and micron if not sub-micron, dimensional accuracies.

Passive mechanical alignment

ULE®'s unique properties make this material also ideal as a substrate for aligning multiple optical components that must remain unaffected by large or even minimal temperature fluctuations to guarantee the optimal performance of sensitive optical systems. In laser cavities, for example, the use of ULE® glass to align mirrors and other optical elements ensures that the optical path length inside the oscillating cavity remains exceptionally stable. This feature is exploited for instance to stabilize lasers used in high-precision interferometric or spectroscopic instruments, both in terms of coherence and output power. There, even the minimal thermal expansion could cause slight misalignments very detrimental to the laser operation, and thus to the system performance [6 - 8, 30]. Similar arguments are used to justify the usage of ULE® substrates inside frequency combs optical generators built for precise frequency measurements and timekeeping [31].

As shown in the example of Figure 3, a glass plate has been machined with precise mechanical features to host various discrete optical elements. The design and the accuracy of these features were such that the optical elements could be placed on the substrate and passively guided into position with a precision of a few microns without the need for any active, optical alignment. Once in place, they were permanently glued, soldered, or welded [33] to the substrate.

In this specific example, a glass fiber aligner was placed in front of a lens that collimates the beam before a diffractive element. The reflected light passed through a Brewster window, before being eventually refocused by a second lens, as schematically shown in Figure 3A. The plate, 70x24 mm in size, was manufactured again by SLE technology with an accuracy that guarantees position and distance deviations measured to be < ± 1.5 µm over lengths up to 30 mm. This approach may not always compete with active alignment procedures [17, 32] but it certainly provides a solid, simpler, and more cost-effective alternative in a multitude of applications for which a few microns of accuracy is sufficient.

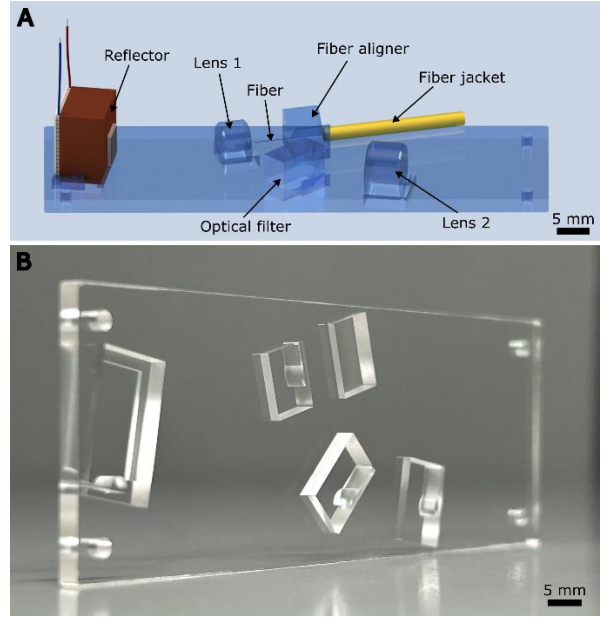

Figure 3. (A) CAD rendering of an optical assembly. (B) ULE substrate fabricated with SLE technology.

## Conclusions

In conclusion, we have demonstrated that ultra-low expansion glass ULE® can be successfully machined using Selective Laser Etching (SLE) technology. The combination of these two elements can lead to extremely stable components, both mechanically and thermally, with almost arbitrary 3D shapes with micron-scale accuracy even for cm-sized objects. Sub-micron surface roughness is also achieved.

The ability to produce such precise and stable components expands the potential applications of ULE® glass, encouraging further innovation and development of highly accurate components to be used in a much broader range of applications, from astronomy to medical and metrological instrumentation, to ultra-stable laser systems and more.

## Acknowledgments

The authors thank Dr. C. Alfieri for fruitful discussions; S. Frisoni, G. Tramezzani, M. Demarta Juillerat, and F.

Comparone for the support in the manufacturing of the components; and M. Gianola for the photography.

**Author Contributions**

**Enrico Casamenti**: Conceptualization (equal), Data curation (equal), Investigation (lead), Methodology (equal), Writing – original draft (equal), Writing – review & editing (lead).
**Alessandro Bruno**: Data curation (equal), Investigation (supporting), Writing – original draft (supporting).
**Pietro Bernasconi**: Conceptualization (equal), Methodology (equal), Supervision (lead), Writing – original draft (equal), Writing – review & editing (supporting).
**Andrea Lovera**: Funding acquisition (lead), Supervision (supporting), Writing – original draft (supporting), Writing – review & editing (supporting).

**References**

[1] C. L. Rathmann, G. H. Mann, and M. E. Nordberg, "A New Ultralow-Expansion, Modified Fused-Silica Glass," *Appl. Opt.*, vol. 7, no. 5, p. 819, May 1968, doi: 10.1364/AO.7.000819.
[2] R. Sabia, M. J. Edwards, R. VanBrocklin, and B. Wells, "Corning 7972 ULE material for segmented and large monolithic mirror blanks," presented at the SPIE Astronomical Telescopes + Instrumentation, E. Atad-Ettedgui, J. Antebi, and D. Lemke, Eds., Orlando, Florida , USA, Jun. 2006, p. 627302. doi: 10.1117/12.672059.
[3] S. T. Gulati and M. J. Edwards, "ULE - Zero expansion, low density, and dimensionally stable material for lightweight optical systems," presented at the Critical Review Collection, Bellingham, United States, Jul. 1997, p. 1028909. doi: 10.1117/12.284710.
[4] N. Jovanovic, R. J. Harris, and N. Cvetojevic, "Astronomical Applications of Multi-Core Fiber Technology," *IEEE J. Select. Topics Quantum Electron.*, vol. 26, no. 4, pp. 1–9, Jul. 2020, doi: 10.1109/JSTQE.2020.2975636.
[5] S. Minardi, R. J. Harris, and L. Labadie, "Astrophotonics: astronomy and modern optics," *Astron Astrophys Rev*, vol. 29, no. 1, p. 6, Dec. 2021, doi: 10.1007/s00159-021-00134-7.
[6] D. Świerad *et al.*, "Ultra-stable clock laser system development towards space applications," *Sci Rep*, vol. 6, no. 1, p. 33973, Sep. 2016, doi: 10.1038/srep33973.
[7] S. Kulkarni *et al.*, "Ultrastable optical components using adjustable commercial mirror mounts anchored in a ULE spacer," *Appl. Opt.*, vol. 59, no. 23, p. 6999, Aug. 2020, doi: 10.1364/AO.395831.
[8] F. Meylahn, N. Knust, and B. Willke, "Stabilized laser system at 1550 nm wavelength for future gravitational-wave detectors," *Phys. Rev. D*, vol. 105, no. 12, p. 122004, Jun. 2022, doi: 10.1103/PhysRevD.105.122004.
[9] T. Asshauer, C. Latz, A. Mirshahi, and C. Rathjen, "Femtosecond lasers for eye surgery applications: historical overview and modern low pulse energy concepts," *Advanced Optical Technologies*, vol. 10, no. 6, pp. 393–408, Dec. 2021, doi: 10.1515/aot-2021-0044.
[10] A. Marcinkevičius *et al.*, "Femtosecond laser-assisted three-dimensional microfabrication in silica," *Opt. Lett.*, vol. 26, no. 5, p. 277, Mar. 2001, doi: 10.1364/OL.26.000277.
[11] Y. Bellouard, A. Said, M. Dugan, and P. Bado, "Fabrication of high-aspect ratio, micro-fluidic channels and tunnels using femtosecond laser pulses and chemical etching," *Opt. Express*, vol. 12, no. 10, p. 2120, 2004, doi: 10.1364/OPEX.12.002120.
[12] V. Maselli, J. R. Grenier, S. Ho, and P. R. Herman, "Femtosecond laser written optofluidic sensor: Bragg grating waveguide evanescent probing of microfluidic channel," *Opt. Express*, vol. 17, no. 14, p. 11719, Jul. 2009, doi: 10.1364/OE.17.011719.
[13] E. Casamenti, S. Pollonghini, and Y. Bellouard, "Few pulses femtosecond laser exposure for high efficiency 3D glass micromachining," *Opt. Express*, vol. 29, no. 22, p. 35054, Oct. 2021, doi: 10.1364/OE.435163.
[14] S. Kiyama, S. Matsuo, S. Hashimoto, and Y. Morihira, "Examination of Etching Agent and Etching Mechanism on Femotosecond Laser Microfabrication of Channels Inside Vitreous Silica Substrates," *J. Phys. Chem. C*, vol. 113, no. 27, pp. 11560–11566, Jul. 2009, doi: 10.1021/jp900915r.
[15] S. I. Nazir and Y. Bellouard, "A Monolithic Gimbal Micro-Mirror Fabricated and Remotely Tuned with a Femtosecond Laser," *Micromachines*, vol. 10, no. 9, p. 611, Sep. 2019, doi: 10.3390/mi10090611.
[16] S. I. Nazir and Y. Bellouard, "Contactless Optical Packaging Concept for Laser to Fiber Coupling," *IEEE Trans. Compon., Packag. Manufact. Technol.*, vol. 11, no. 6, pp. 1035–1043, Jun. 2021, doi: 10.1109/TCPMT.2021.3080513.
[17] A. Delgoffe, S. Nazir, S. Hakobyan, C. Hönninger, and Y. Bellouard, "All-glass miniature GHz repetition rate femtosecond laser cavity," *Optica*, vol. 10, no. 10, p. 1269, Oct. 2023, doi: 10.1364/OPTICA.496503.
[18] Y. Bellouard, A. A. Said, and P. Bado, "Integrating optics and micro-mechanics in a single substrate: a step toward monolithic integration in fused silica," *Opt. Express*, vol. 13, no. 17, p. 6635, Aug. 2005, doi: 10.1364/OPEX.13.006635.

[19] R. Osellame, V. Maselli, R. M. Vazquez, R. Ramponi, and G. Cerullo, "Integration of optical waveguides and microfluidic channels both fabricated by femtosecond laser irradiation," *Appl. Phys. Lett.*, vol. 90, no. 23, p. 231118, Jun. 2007, doi: 10.1063/1.2747194.

[20] M. Haque and P. R. Herman, "Chemical-assisted femtosecond laser writing of optical resonator arrays: Chemical-assisted femtosecond laser writing of optical resonator arrays," *Laser & Photonics Reviews*, vol. 9, no. 6, pp. 656–665, Nov. 2015, doi: 10.1002/lpor.201500062.

[21] E. Casamenti, T. Yang, P. Vlugter, and Y. Bellouard, "Vibration monitoring based on optical sensing of mechanical nonlinearities in glass suspended waveguides," *Optics Express*, vol. 29, no. 7, pp. 10853–10862, 2021, doi: 10.1364/OE.414191.

[22] L. Bonnefoy, C. Baur, Y. Bellouard, and S. Henein, "Bistable Puncturing Tool for Retinal Vein Cannulation," 2020.

[23] E. Casamenti, C. Alfieri, R. Ferrini, and A. Lovera, "Laser-based micro-machining of high precision glass ferrules for 2D fiber arrays," in *Technical Digest Series (Optica Publishing Group, 2023)*, Optica Publishing Group, 2023, pp. AM4R-4. doi: 10.1364/CLEO_AT.2023.AM4R.4.

[24] S. Richter *et al.*, "Laser induced nanogratings beyond fused silica - periodic nanostructures in borosilicate glasses and ULE$^{TM}$," *Opt. Mater. Express*, vol. 3, no. 8, p. 1161, Aug. 2013, doi: 10.1364/OME.3.001161.

[25] S. Richter *et al.*, "Ultrashort pulse induced modifications in ULE - from nanograting formation to laser darkening," *Opt. Mater. Express*, vol. 5, no. 8, p. 1834, Aug. 2015, doi: 10.1364/OME.5.001834.

[26] M. Lancry *et al.*, "Nanogratings formation in multicomponent silicate glasses," *Appl. Phys. B*, vol. 122, no. 3, p. 66, Mar. 2016, doi: 10.1007/s00340-016-6337-8.

[27] I. Efthimiopoulos *et al.*, "Femtosecond laser-induced transformations in ultra-low expansion glass: Microstructure and local density variations by vibrational spectroscopy," *Journal of Applied Physics*, vol. 123, no. 23, p. 233105, Jun. 2018, doi: 10.1063/1.5030687.

[28] M. Cavillon, Y. Wang, B. Poumellec, F. Brisset, and M. Lancry, "Erasure of nanopores in silicate glasses induced by femtosecond laser irradiation in the Type II regime," *Appl. Phys. A*, vol. 126, no. 11, p. 876, Nov. 2020, doi: 10.1007/s00339-020-04062-8.

[29] G. Torun, "Femtosecond laser-induced modifications and self-organization in complex glass systems," EPFL, 2023. [Online]. Available: https://doi.org/10.5075/epfl-thesis-10056

[30] C. G. E. Alfieri *et al.*, "Optical efficiency and gain dynamics of modelocked semiconductor disk lasers," *Opt. Express*, vol. 25, no. 6, p. 6402, Mar. 2017, doi: 10.1364/OE.25.006402.

[31] T. W. Hänsch *et al.*, "Precision spectroscopy of hydrogen and femtosecond laser frequency combs," *Phil. Trans. R. Soc. A.*, vol. 363, no. 1834, pp. 2155–2163, Sep. 2005, doi: 10.1098/rsta.2005.1639.

[32] B. Pezeshki *et al.*, "High Performance MEMS-Based Micro-Optic Assembly for Multi-Lane Transceivers," *J. Lightwave Technol.*, vol. 32, no. 16, pp. 2796–2799, Aug. 2014.

[33] S. Richter, S. Döring, T. Peschel, R. Eberhardt, S. Nolte, and A. Tünnermann, "Breaking stress of glass welded with femtosecond laser pulses at high repetition rates," presented at the SPIE LASE, A. Heisterkamp, J. Neev, and S. Nolte, Eds., San Francisco, California, USA, Feb. 2011, p. 79250P. doi: 10.1117/12.874529.